\documentclass[twocolumn,aps,showpacs,prb,lettersize]{revtex4}
\usepackage{graphicx}
\usepackage{dcolumn}
\usepackage{bm}
\usepackage{amsfonts}
\usepackage{latexsym}
\usepackage{amssymb}
\usepackage{layout}

\def \vf{${\bf v}_{\rm F}$}
\def \k{${\bf k}$}
\def \kd{${\bf k'}$}
\def \mfp{${\bf l}$}

\def \kkd{$\mathbb{P}_{{\bf kk'}}$}

\def \kf{${\bf k}_{\rm F}$}
\def \tl{Tl2201}
\begin{document}

\title{Angle dependent magnetoresistance measurements in Tl$_2$Ba$_2$CuO$_{6+\delta}$ and the need for anisotropic scattering}

\author{J. G. Analytis, M. Abdel-Jawad, L. Balicas$^\dag$, M. M. J. French and N. E. Hussey}

\affiliation{H. H. Wills Physics Laboratory, University of Bristol, Tyndall Avenue, Bristol, BS8 1TL, UK}
\affiliation{$^\dag$National High Magnetic Field Laboratory, Florida State University, Tallahassee, FL-32306, USA}

\begin{abstract} 
The angle-dependent interlayer magnetoresistance of overdoped Tl$_2$Ba$_2$CuO$_{6+\delta}$ has been measured in high
magnetic fields up to 45 Tesla. A conventional Boltzmann transport analysis with no basal-plane anisotropy in the
cyclotron frequency $\omega_c$ or transport lifetime $\tau$ is shown to be inadequate for explaining the data. We
describe in detail how the analysis can be modified to incorporate in-plane anisotropy in these two key quantities and
extract the degree of anisotropy for each by assuming a simple four-fold symmetry. While anisotropy in $\omega_c$ and
other Fermi surface parameters may improve the fit, we demonstrate that the most important anisotropy is that in the
transport lifetime, thus confirming its role in the physics of overdoped superconducting cuprates.
\end{abstract}



\maketitle
\section{Introduction}
There are many routes to investigating the mechanism of high-temperature superconductivity and naively one might expect
the normal state to be the simplest. Despite concerted experimental effort however, \cite{husseyrev} the normal state
properties of cuprates remain a profound theoretical challenge. \cite{zaanen06} Indeed, even from the earliest
transport measurements in these compounds it was clear that the normal state was far from conventional.
\cite{gurvitchfiory87} Arguably the most remarkable phenomena are the distinct power laws of the in-plane resistivity
$\rho_{ab}$ and inverse Hall angle ${\rm cot}\Theta_{\rm H}$ temperature dependences. In optimally doped
YBa$_2$Cu$_3$O$_{7-\delta}$ (YBCO) and La$_{2-x}$Sr$_x$CuO$_4$ (LSCO) for example, $\rho_{ab}(T)$ varies linearly with
temperature over a wide temperature range, whereas ${\rm cot}\Theta_{\rm H}$ maintains a strong $T^2$ dependence.
\cite{chien91,hwang94} In other words, it is as if these materials exhibit distinct scattering mechanisms which are
separately manifested according to the experimental probe being considered. Anderson coined the phrase \lq lifetime
separation' to describe this anomalous behavior and today its interpretation remains one of the greatest obstacles to
the development of a coherent description of the normal state quasiparticle dynamics in high-$T_c$ cuprates.

Three contrasting approaches dominate the current thinking on the transport problem in cuprates; Anderson's
two-lifetime picture,\cite{anderson} marginal Fermi-liquid (MFL) phenomenology \cite{varma89} and models based on
fermionic quasiparticles that invoke specific (anisotropic) scattering mechanisms within the basal
plane.\cite{carrington92, monthouxpines92, castellani95, ioffemillis98, hussey03b} In the two-lifetime approach,
scattering processes involving momentum transfer perpendicular and parallel to the Fermi surface are governed by
independent transport and Hall scattering rates 1/$\tau_{tr}$ and 1/$\tau_{\rm H}$ with different $T$-dependencies. The
proponents of the MFL hypothesis assume a single $T$-linear scattering rate which naturally accounts for
$\rho_{ab}(T)$, but introduce an unconventional expansion in the magnetotransport response whereby the Hall angle, for
example, is given by the square of the transport lifetime. \cite{varmaabrahams01} This anomalous expansion is
attributed to anisotropy in the (elastic) impurity scattering rate, possibly due to small-angle scattering off
impurities located away from the CuO$_2$ plane. \cite{varmaabrahams01}

Attempts to explain the anomalous behavior of $\rho_{ab}(T)$ and ${\rm cot}\Theta_{\rm H}(T)$ in cuprates within a
Fermi-liquid (FL) approach have centered around the assumption of a (single) {\it inelastic} scattering rate that is
strongly dependent on the quasiparticle wave-number ${\bf k}$. This anisotropy can arise either due to anisotropic
electron-electron (possibly Umklapp) scattering \cite{hussey03b} or coupling to a singular bosonic mode, be that of
spin, \cite{carrington92, monthouxpines92} charge \cite{castellani95} or $d$-wave superconducting fluctuations.
\cite{ioffemillis98} Generating a clear separation of lifetimes within these single-lifetime scenarios however requires
a subtle balancing act between different regions in {\bf k}-space with distinct $T$-dependencies. \cite{sandeman01}

In order to test these various proposals and to proceed towards a theoretical consensus, information on the momentum
({\bf k}) and energy ($\omega$ or $T$) dependence of the transport lifetime $\tau$ at or near the Fermi level
$\epsilon_F$ is urgently required. This is a non-trivial exercise however since the transport coefficients themselves
are angle-averaged quantities involving differently weighted integrations around the Fermi surface (FS). Whilst angle
resolved photoemission spectroscopy (ARPES) can probe directly the in-plane quasi-particle lifetime via the imaginary
part of the self-energy Im$\Sigma$({\bf k},$\omega$), its relevance to DC transport is still unclear. \cite{majed07}
Moreover, there remains some dispute as to the correct form of Im$\Sigma$({\bf k},$\omega$) even for samples with
nominally the same composition.\cite{kordyuk04, kaminski05}

Measurements of interlayer magnetoresistance as a function of angle have yielded important information about the FS
topology (size and shape) in a variety of layered metals including organic conductors\cite{kartsovnik04} and
quasi-two-dimensional (q2D) oxides. \cite{bergemann03, hussey03a, balicas05} In a recent paper, we showed that this
technique could be developed to extract information on the scattering rate anisotropy and applied the technique to
overdoped Tl$_2$Ba$_2$CuO$_{6+\delta}$ (\tl). \cite{majed06} In the present paper, we present a more thorough and
detailed analysis of our angle-dependent magnetoresistance (ADMR) measurements on overdoped \tl, focusing in particular
on the procedure used to fit ADMR and show how this analysis fails at higher temperatures unless one includes such
anisotropy in $\tau$. We progressively introduce anisotropy into the formalism, and explore the effects of this both in
the cyclotron frequency $\omega_c$ and the transport lifetime $\tau$. The approach presented here is similar to that
described recently by Kennett and McKenzie who derived a generalized expression for ADMR in layered metals with
basal-plane anisotropy. \cite{kennett06} In this paper, we focus on issues pertinent to \tl, the importance of each
parameter in fitting the ADMR signal and their interdependence, and the issue of sample misalignment. Although it is
difficult to isolate anisotropy in one from anisotropy in the other, the strong temperature evolution of the ADMR
signal (and subsequent measurements of its doping dependence \cite{majed07}) suggests that the dominant anisotropy is
in $\tau$ and not $\omega_c$. The paper is set out as follows. Section \ref{threedfs} describes the FS parameterization
of \tl~and the necessary symmetry considerations with respect to the ADMR analysis. Section \ref{expt} briefly
describes the ADMR experiment itself. The Boltzmann formalism and the resulting analysis is described in Section
\ref{sims} for the cases where the parameters $\omega_c$ and $\tau$ are both isotropic and anisotropic (within the
basal-plane). Our conclusions are presented in Section \ref{conc}.

\section{The three-dimensional Fermi surface of Tl$_2$Ba$_2$CuO$_{6+\delta}$}
\label{threedfs}

\begin{figure}[tp]
\includegraphics[width=7cm,angle=0]{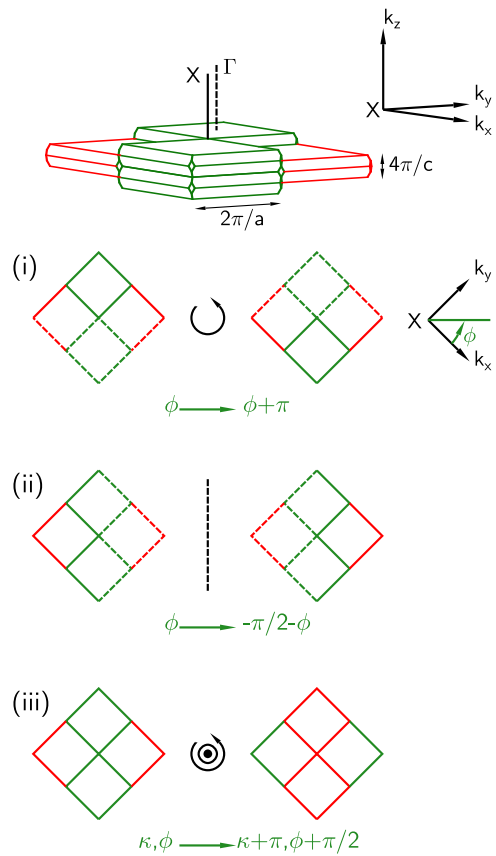}
\caption{(Color online) The Brillouin zone stacking for Tl$_2$Ba$_2$CuO$_{6+\delta}$. There is inversion symmetry about
the $k_x-k_y$ plane: $\kappa,\phi\rightarrow-\kappa,\phi$ and three further symmetries about the axis along the
$X$-line. As described in the text they are (i) the two fold symmetry $\phi\rightarrow\phi+\pi$, (ii) the mirror
symmetry $\phi\rightarrow-\pi/2-\phi$ and finally (iii) the screw symmetry involving a translation in the $k_z$
direction and a rotation: $\kappa,\phi\rightarrow\kappa+\pi,\phi+\pi/2$.} \label{bztl}
\end{figure}

\begin{figure}[htp]
\includegraphics[width=8cm,angle=0]{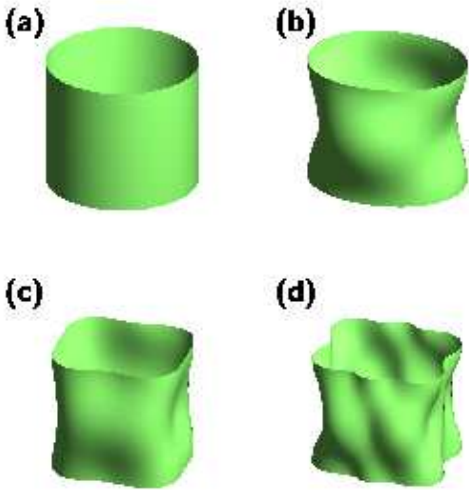}
\caption{(Color online) Quasi-2D Fermi surfaces described by Eq. \ref{outplane} with progressive inclusion of
cylindrical harmonics (a) $k_{00}$, (b) $k_{21}$, (c) $k_{04}$ and (d) $k_{61}$ and $k_{101}$.} \label{tlfs}
\end{figure}

The present FS parameterization is identical to that used previously \cite{hussey03a, majed06, kennett06} and so shall
be described here only briefly. The interested reader is referred to [Ref. \cite{bergemann03}] which details a similar
parameterization of the $\alpha$ sheet of ${\rm Sr_2RuO_4}$. Being extended in the $k_z$ direction, the quasiparticle
dispersion contains a finite (though small $\sim$ meV) transfer integral $t_{\perp}(\phi,k_z)$, where $\phi$ is the
azimuthal angle in the $k_x-k_y$ plane. The parameter $t_{\perp}(\phi,k_z)$ is anisotropic in the plane and the Fermi
wave-vector $k_F$ is therefore modulated by both the in-plane dispersion and by $t_{\perp}(\phi,k_z)$. The clearest way
to express this is by expanding $k_F$ into cylindrical harmonics\cite{bergemann03,hussey03a}

\begin{equation}
  k_{\rm F}(\phi,\kappa) = \displaystyle\sum_{m,n = 0}k_{mn}^{\footnotesize\left\{\begin{array}{l}{\rm c}\\{\rm s}\end{array}\right\}\left\{\begin{array}{l}{\rm c}\\{\rm s}\end{array}\right\}}\left\{\begin{array}{l}{\rm cos}\\{\rm sin}\end{array}\right\}n\kappa\times\left\{\begin{array}{l}{\rm cos}\\{\rm sin}\end{array}\right\}m\phi.
\label{outplane0}
\end{equation}

where $\{^{\rm c}_{\rm s}\}$ denote coefficients corresponding to cosine and sine terms, $\kappa=k_zc/2$ and
$c=23.2\mbox{\AA}$ is the interlayer dimension of the unit cell.  The symmetry of the Brillouin zone limits the number
of parameters of interest and a pictorial illustration of this is shown in Figure \ref{bztl}.  In the $k_z$ direction,
inversion symmetry $\kappa\rightarrow-\kappa$ requires that the only terms containing $\kappa$ are cosines. Three
further symmetries restrict the parameterization: (i) the two-fold rotational symmetry $\phi\rightarrow\phi+\pi$, (ii)
the mirror plane $\phi\rightarrow-\pi/2-\phi$ and (iii) the screw symmetry
$\kappa,\phi\rightarrow\kappa+\pi,\phi+\pi/2$.  The transformations differ from those in Ref. \cite{bergemann03}
because of a different choice in coordinate axes, but the operations are identical. The first symmetry requires that
all $m$ be even. The next symmetry requires that all cosine terms have $m\,{\rm mod}4\equiv0$ and all sine terms have
$m\,{\rm mod}4\equiv2$.  For example, ${\rm cos4\phi}={\rm cos}4(-\pi/2-\phi)$ whereas ${\rm cos2\phi}=-{\rm
cos}2(-\pi/2-\phi)$. The reverse is true for the sine terms. The final symmetry requires that all of the cosine terms
be accompanied by $n$ that are even and the sine terms be accompanied by any $n$ that are odd. For example, ${\rm cos
\kappa}{\rm sin2\phi}={\rm cos(\kappa+\pi)}{\rm sin}2(\pi/2+\phi)$, but ${\rm cos 2\kappa}{\rm sin2\phi}=-{\rm
cos2(\kappa+\pi)}{\rm sin}2(\pi/2+\phi)$. The converse is of course true for the cosine terms that have $m\,{\rm
mod}4\equiv0$. Eq. \ref{outplane0} can thus be simplified to

\begin{equation}
\begin{array}{lll}
k_{\rm F}(\phi,\kappa)& =& \displaystyle\tiny \sum_{\begin{array}{l}m,n = 0\\ m\,{\rm mod}\,4 = 0\\n\, even\end{array}}k_{mn}{\rm cos}(n\kappa){\rm cos}(m\phi)\\
          &&\normalsize \displaystyle\tiny+\sum_{\begin{array}{l}m,n = 0\\ m\,{\rm mod}\,4 = 2\\n\, odd\end{array}}k_{mn}{\rm cos}(n\kappa){\rm sin}(m\phi).\\
\end{array}
\label{outplane}
\end{equation}

\begin{figure}[htp]
\includegraphics[width=7cm,angle=0]{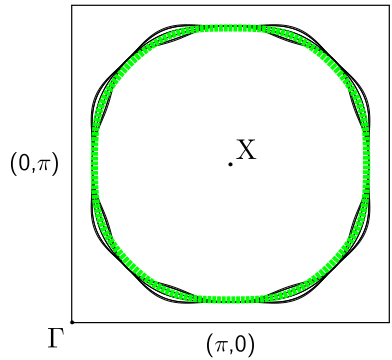}
\caption{(Color online) The projection of the 3D dispersion for overdoped \tl\, as determined by ADMR (thin black
lines), plotted over the ARPES (thick green line) results for a compound with a nominally similar doping.\cite{plate05}
The warping of the transfer integral is exaggerated by approximately 100 times for clarity.} \label{arpescomp}
\end{figure}

We have shown previously that the minimum number of parameters required to fit the data that simultaneously satisfy
these symmetry constraints are $k_{00},k_{04},k_{21},k_{61}$ and $k_{101}$.\cite{hussey03a} Figure \ref{tlfs} shows the
warping created by progressive inclusion of these parameters, beginning with a dispersionless ($t_{\perp}=0$) isotropic
FS. Eq. \ref{outplane} has exact four-fold symmetry though the modulation $t_{\perp}$ gives rise to eight highly
symmetric points where the transfer integral vanishes, as predicted by band-structure calculations. \cite{andersen95}
Figure \ref{arpescomp} shows the projection of the three-dimensional FS as deduced by ADMR \cite{hussey03a} overlaid on
that determined by ARPES \cite{plate05} (see Eq. \ref{inplane}; this curve corresponds to the nominal doping level of
this crystal). The agreement is very good, but most importantly the two experiments, to a good approximation, share the
eight points of high symmetry. For ease of computation, the Fermi surface can be described by

\begin{equation}
\begin{array}{lll}
\epsilon_{\rm F}(\kappa,\phi)&=&{\hbar^2\over 2m}k^{\parallel2}_{\rm F}(\phi)-2t_{\perp}{\rm cos\kappa}\\
&&\\
&&\times{a\over\pi}(k_{21}{\rm sin}2\phi+k_{61}{\rm sin}6\phi+k_{101}{\rm sin}10\phi),
\end{array}
\label{tldisp}
\end{equation}
where $k^{\parallel}_{\rm F}(\phi)=k_{00}+k_{04}{\rm cos}(4\phi)$ and $a=3.866\mbox{\AA}$ is the
in-plane lattice parameter.

\section{Experimental}
\label{expt}

\begin{figure}[tph]
\includegraphics[width=6.3cm,angle=0]{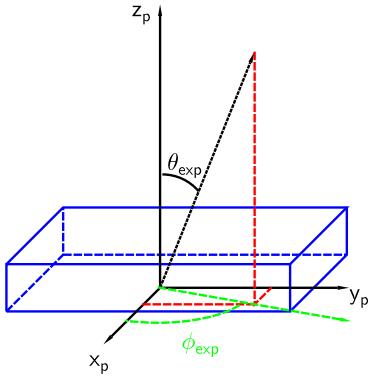}
\caption{(Color online) Diagram describing the ADMR experimental technique, whereby the sample (schematically shown in
blue) is rotated by an azimuthal angle $\phi_{\rm exp}$ with respect to the laboratory frame ($x_p,y_p,z_p$), and then
data are continuously taken as a function of polar angle $\theta_{\rm exp}$. In the actual experiment, the
crystallographic $(x_c,y_c)$ plane does not lie exactly in the laboratory frame due to a slight misalignment and the
corresponding polar angle $\theta_{\rm crys}\equiv\theta$ (taken as the angle between the field direction and the
normal to the plane on the sample) differs from $\theta_{\rm exp}$. Moreover, the azimuthal $\phi_{\rm crys}$ may
change as a function of $\theta_{\rm exp}$ as explained in Appendix \ref{appasym}. } \label{rotation}
\end{figure}

\tl\, is the most suitable cuprate system for ADMR studies due to its single Fermi sheet, \cite{peets06} its strong
two-dimensionality, \cite{hussey94} its low residual resistivity \cite{mackenzie96, hussey96, proust02} and
accessibility to the whole overdoped region of the cuprate phase diagram. \cite{kubo91} Single crystals were fabricated
using a self-flux method in alumina crucibles. \cite{tyler97} As-grown crystals are naturally overdoped and the doping
level (and therefore the desired $T_c$) is set by annealing in oxygen, argon or in vacuum. \cite{tyler97} The crystal
used in this study (300$\mu$m x 150$\mu$m x 20$\mu$m) was annealed in oxygen at 600K for 200min resulting in
$T_c\approx 17$K. Electrical contacts were attached using Dupont 6838 silver paste in a quasi-Montgomery 4-wire
configuration. ADMR measurements were performed at 45T in the hybrid magnet at the National High Magnetic Field
Laboratory, Tallahassee, Florida using a probe with a two-axis rotator. Initially, the platform on which the sample was
mounted was rotated by an azimuthal angle $\phi_{\rm exp}$ and then the interplane resistivity $\rho_{zz}$ was measured
as the polar angle $\theta_{\rm exp}$ was swept at constant temperature and constant field (see Figure \ref{rotation}).

\section{Fitting of the angle-dependent magnetoresistance in \tl}
\label{sims}

In this section we review how the Boltzmann transport equation can be used to fit the ADMR data. We begin with the
simplest case whereby both $\omega_c$ and $\tau$ are isotropic before going on to discuss the more general case in
which both parameters are anisotropic within the conducting plane.

\subsection{Isotropic $\tau$ and $\omega_c$}
\label{isoall}

\begin{figure*}[tp]
\includegraphics[width=10.5cm,angle=0]{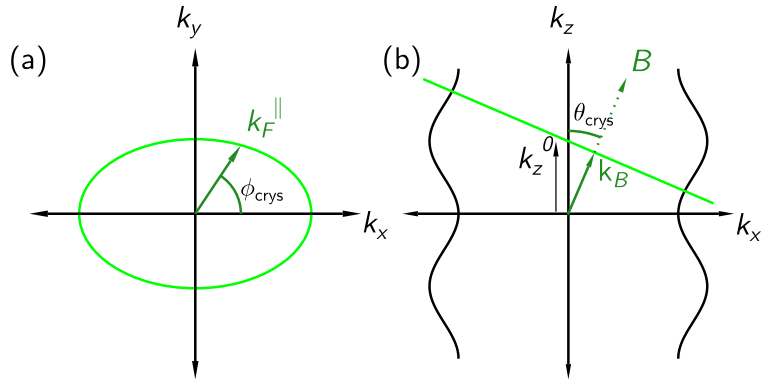}
\caption{(Color online) Schematic representation of a single quasiparticle orbit with an oriented magnetic field. Panel
(a) gives the projected view of the Fermi surface, defining the azimuthal angle $\phi_{\rm crys}$, while panel (b)
gives the definition of the parameters ${\rm {\bf k}}_B$, $k_z^0$ and $\theta_{\rm crys}$.} \label{yamajifs}
\end{figure*}

In the presence of a magnetic field a quasiparticle traverses the FS following the contours defined by the dispersion.
During this journey the quasiparticle will gain velocity from the electric field until it encounters a scattering
event, after which it begins its journey again. As the angle of the field with respect to the crystal axes is adjusted,
the quasiparticle will traverse different orbits and the average velocity in the direction of the current can vary
dramatically. This picture is formalized in the Chambers' tube integral, which is the solution to the Boltzmann
transport equation in the relaxation-time approximation

\begin{equation}
\sigma_{ij}={e^2\over 4\pi^3\hbar^2}\int_{FS} d{\bf k}\left({\partial f_{\rm{\bf k}}\over\partial {\rm{\bf
k}}}\right)v_i({\bf k},0)\int_{-\infty}^0v_j({\bf k},t)e^{t/\tau}dt,
\end{equation}

where $f_{\rm {\bf k}}$ is the mean occupation of state ${\rm {\bf k}}$ and $\tau$ is assumed to be independent of \k\,
(or equivalently, isotropic in the azimuthal angle $\phi$). The Chambers formula can be used in a situation where both
closed and open orbits are present.\cite{blundell97, goddard04} For our particular interest only closed orbits are
involved (the FS is q-2D)\cite{hussey03a, plate05} and we are able to use the simpler Shockley-Chambers tube integral.
Furthermore it is easier to use cylindrical coordinates, in line with our description of $k_{\rm F}(\kappa,\phi)$. The
interplane conductivity is then given by

\begin{equation}
\begin{array}{lll}
\sigma_{zz}& =& \displaystyle{e^2\over4\pi^3\hbar^2}\int d\varepsilon \left({-\partial f^0\over\partial\varepsilon}\right)\int d{\bf k}_B\\
&&\\
&&\displaystyle\times\int_0^{2\pi} d\phi {v_z(\phi,{\bf k}_B,\varepsilon)\over \omega_c}\int_0^\infty d\phi' {v_z(\phi-\phi',{\bf k}_B,\varepsilon)\over \omega_c}\\
&&\\
&& \displaystyle\,\times{\rm e}^{-\phi/\omega_c\tau},
\end{array}
\label{condi0}
\end{equation}

where ${\bf k}_B$ is the reciprocal space direction parallel to the magnetic field ${\bf H}$, $d\phi=\omega_c dt$ and
$\omega_c$ is considered isotropic. In order to use Eq. \ref{condi0} to analyse the ADMR data, we follow Yamaji
\cite{yamaji} and define a vector $k_z^0$ as shown in Figure \ref{yamajifs}(b). The projection of the magnetic field on
the azimuthal plane relative to the $k_x$-axis (corresponding to the Cu-O-Cu bond direction) is labelled $\phi_{\rm
crys}$ (see Figure \ref{yamajifs} (a)). Each orbital plane is then defined by three parameters: the polar angle
$\theta_{\rm crys}\equiv\theta$, the azimuthal angle $\phi_{\rm crys}$ and $k_z^0$. The former two are determined
during the experiment whereas the latter is an integration variable in the fitting procedure described below.
\cite{asym} The intersection of this plane with the FS gives the path of the quasiparticle in reciprocal space. This
plane is given by the equation

\begin{equation}
k_x{\rm sin\theta}+k_z{\rm cos\theta} = \vert{\bf k}_B\vert = k_z^0{\rm cos\theta}.
\end{equation}

This is a convenient notation because $k_z$ can be uniquely described in terms of $k_z^0$ and the projection of the
Fermi wave-vector onto the azimuthal plane as the quasiparticle traverses an orbit, ${\bf k}_F^{\parallel}(\phi)$. In
summary

\begin{equation}
k_z = \vert{\bf k}_B\vert = k_z^0-k_F^{\parallel}(\phi){\rm cos}(\phi-\phi_{\rm crys}){\rm tan\theta}.
\label{kzf}
\end{equation}

We make the replacement $\varepsilon\rightarrow\varepsilon_F$ in Eq. \ref{condi0} by approximating $\partial
f^0/\partial\varepsilon\rightarrow\delta(\varepsilon-\varepsilon_F)$ for $kT\ll\varepsilon_F$. The periodicity of
$v_z(\phi)$ and $v_z(\phi-\phi')$ in $\phi$ and $\phi-\phi'$ respectively is of some computational benefit. Taking the
Fourier transform of $v_z$ \cite{majedthesis,yagi} and writing it as a Fourier sum gives

\begin{equation}
\begin{array}{l}
v_z(\phi)=\displaystyle a_0+\sum_0^{\infty}a_n{\rm cos }\,n\phi+b_n{\rm sin} \,n\phi,\\
v_z(\phi-\phi')=\displaystyle c_0+\sum_0^{\infty}c_n{\rm cos }\,n(\phi-\phi')+d_n{\rm sin} \,n(\phi-\phi'),
\end{array}
\label{fftvi}
\end{equation}

where $a_n,b_n,c_n$ and $d_n$ are Fourier coefficients. Using a Laplace transform and after a few algebraic
manipulations, the conductivity is finally given by \cite{majedthesis,yagi}

\begin{equation}
\begin{array}{lll}
\sigma_{zz}&=&\displaystyle{e^3\tau_0B{\rm cos}(\theta)\over 2\pi^2\hbar^2}\int dk_z^0\omega_0^{-1}\\
&&\\
&& \displaystyle\times\left[a_0c_0+{1\over2}\sum_{n=1}^{\infty}\left\{{a_nc_n+b_nd_n\over1+(\omega_0\tau_0)^2}-\right.\right.\\
&&\\
&&\displaystyle\left.\left.{(a_nd_n-b_nc_n)\omega_0\tau_0n\over1+(\omega_0\tau_0)^2}\right\}\right],
\end{array}
\label{fftsigA}
\end{equation}

where $\omega_c=\omega_0$ and $\tau=\tau_0$ to emphasize that these parameters are isotropic. Eq. \ref{fftsigA} is used
to calculate the resistivity in the transverse direction $\rho_{zz}$ by taking the inverse of $\sigma_{zz}$, which is
correct to a good approximation due to the large anisotropy of the in-plane and interplane resistivity. Because
parameters such as the effective mass $m^*$ are not well known, it is usual practice to simulate the relative change in
magnetoresistivity $\Delta\rho_{zz}/\rho_{zz0}$, where $\rho_{zz0}$ is the interplane resistivity at zero field, rather
than $\rho_{zz}$ directly. This normalization procedure means that the warping parameters in the $k_z$ direction can
only be determined as {\it ratios}. In other words, the ADMR can be used to obtain values for $k_{61}/k_{21}$ and
$k_{101}/k_{21}$ but not $k_{21},k_{61}$ or $k_{101}$ directly.

The parameters we wish to determine therefore are $k_{00}, k_{04}, k_{61}/ k_{21}, k_{101}/k_{21}$ and $\omega_0\tau_0$
(the cyclotron frequency and the scattering time always appear as a product in the sum of Eq. \ref{fftsigA} and thus
behave as a single parameter). In a number of earlier studies on different \tl\, crystals, in which all parameters were
allowed to vary, a consistent set of FS parameters were obtained. \cite{hussey03a, majed06, majed07} This enables us to
refine our parameterization and minimize the number of free parameters without losing confidence in their relative
magnitudes. We fix $k_{00}$ for example by first obtaining the doping level $p$ using the universal phenomenological
relation \cite{tallon2} between $p$ and the critical temperature $T_c$

\begin{equation}
{T_c(p)\over T_c^{max}}\approx1-82.6(p-0.16)^2,
\end{equation}
then adopting the simple hole-counting procedure,

\begin{equation}
(\pi k_{00}^2)/(2\pi/a)^2 = (1+p)/2.
\end{equation}
Our next simplifying assumption is that $t_{\perp}$($\phi$) vanishes at eight symmetry points on the FS (see Figure
\ref{arpescomp}) as expected from band structure calculations \cite{andersen95} and revealed by earlier ADMR
measurements.\cite{hussey03a} For this to be the case, we require

\begin{equation}
1-{k_{61}\over k_{21}}+{k_{101}\over k_{21}}=0,
\end{equation}
which fixes $k_{101}/ k_{21}$ to whatever value $k_{61}/ k_{21}$ is given. Hence, only three parameters $k_{04}$,
$k_{61}/ k_{21}$ and the product $\omega_0\tau_0$, are used to fit {\it simultaneously} five polar angle sweeps at
different azimuthal angles (in other words, the data are treated as a single data set, not five separate curves). It is
important to realize that these constraints could be relaxed without a significant effect on the other key parameters.

The fitting procedure begins by evaluating $\sigma_{zz}$ for a given polar angle $\theta_{\rm crys}$ and azimuthal
angle $\phi_{\rm crys}$. The $c$-axis velocity is evaluated $v_z$ = $\hbar^{-1}\partial \varepsilon ({\bf
k})/\partial{\it k_z}$ as a function of $\phi$ for a given $k_z^0$, where $\varepsilon$ is defined by Eq. \ref{tldisp}.
The $k_z$ dependence is determined by Eq. \ref{kzf} and substituted into $v_{z}$. For the given $k_z^0$, the Fourier
transform is taken and the sum in Eq. \ref{fftsigA} is evaluated. This is then integrated over $k_z^0$ across the whole
Brillouin zone and the result inverted to give $\rho_{zz}(\theta_{\rm crys},\phi_{\rm crys})$. This is calculated for
all $\theta_{\rm crys}$ and $\phi_{\rm crys}$ in a single data set. This process is repeated for different parameter
values until a best fit is achieved using standard minimization procedures.

\begin{figure*}[tph]
\includegraphics[width=16.88cm,angle=0]{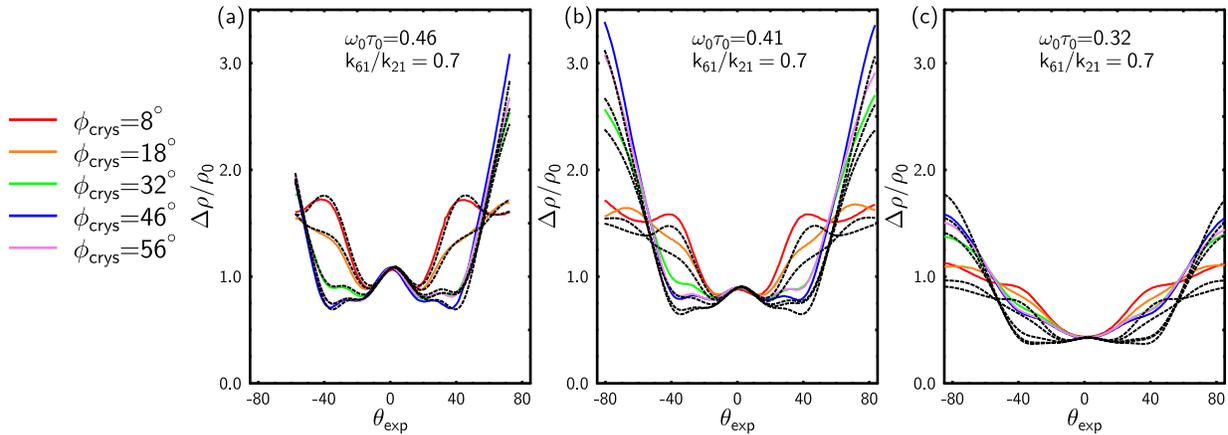}
\caption{(Color online) The solid lines are $c$-axis magnetoresistivity data $\Delta\rho_{zz}=\rho_{zz}(H)-\rho_{zz0}$,
normalized to the zero field resistivity $\rho_{zz0}$, taken at different azimuthal rotations ($\phi_{\rm crys}=
8^\circ$ (red), $\phi_{\rm crys}= 18^\circ$ (orange), $\phi_{\rm crys}= 32^\circ$ (green), $\phi_{\rm crys}= 46^\circ$
(blue), $\phi_{\rm crys}= 56^\circ$ (violet) relative to the Cu-O-Cu bond direction) at three different temperatures
(a) 4.2 K, (b) 14 K and (c) 50 K. The azimuthal angles given here strictly apply only to $\theta_{\rm exp}=\pm 90$ due
to misalignment of the crystal with respect to the platform axes (see Appendix \ref{appasym} for details). The black
dashed lines are the best least-squares fits obtained assuming that $\omega_c\tau$ (= $\omega_0\tau_0$) is independent
of $\phi$ and that the parameters $k_{04}$ and $k_{61}/k_{21}$ are fixed to their values at $T=4.2 $K. Thus only
$\omega_0\tau_0$ is allowed to vary with temperature.}
 \label{isofit}
\end{figure*}

The solid lines in Figure \ref{isofit}(a) are ADMR data taken at $T=4.2\,$K and $\mu_0H$ = 45 Tesla, normalized to the
zero field resistivity value $\rho_{zz0}$. Each color represents a different azimuthal angle at which the individual
polar ADMR sweeps were taken. Despite the fact that $\omega_0\tau_0$ is less than 0.5 in this sample, the variations in
the $c$-axis magnetoresistance are significant, both with azimuthal and polar angle, thus tightly constraining our
parameterization. Note that these data were obtained on a different crystal to those reported in Refs. \cite{hussey03a,
majed06} though the resulting parameterization ($k_{00}$ = 0.729$\AA^{-1}$, $k_{04}$ = -0.022$\AA^{-1}$,
$k_{61}/k_{21}$ = 0.7) is very similar. The best least-squares fits to Eq. \ref{fftsigA} are shown as black dashed
lines and appear quite adequate for the full range of azimuthal and polar angles studied. (Data at larger angles were
not taken at this temperature in order to avoid the large torque forces that accompany a transition to the
superconducting state, which occurs here when $H_{c2}(\theta)$ surpasses 45 T.)

Corresponding data and fits for $T=14 $K and $T= 50 $K are shown in
panels (b) and (c) respectively. For the fits at higher temperatures
(where a larger angular range can be swept), all FS parameters are
fixed to their 4.2 K values and only the product $\omega_0\tau_0$ is
allowed to vary. The fits rapidly deteriorate as the temperature is
raised and are clearly no longer a reliable representation of the
data. In fact, even if we allow $k_{00}$,$k_{04}$ and $k_{61}/k_{21}$
to vary with temperature, the fits do not significantly
improve. Furthermore, if $k_{00}$ is allowed to be a free parameter,
the fitting procedure tends to minimize at values where the Fermi
surface is larger that the first Brillouin zone, which is clearly
unphysical. In response to this failing, we abandon our naive picture
of isotropic $\omega_c$ and $\tau$ and proceed to incorporate
anisotropy into the formalism.

\subsection{Isotropic $\tau$ and anisotropic $\omega_c$}
\label{aniomega}

To illustrate how significant anisotropy in $\omega_c$ can be, we consider first the most elementary tight-binding
description of an isotropic square 2D lattice. The dispersion of such a system can be described by the equation

\begin{equation}
\varepsilon ({\bf k})-\varepsilon_0 = - 2t[{\rm cos}(k_xa)+{\rm
cos}(k_ya)],
\end{equation}
where $\varepsilon-\varepsilon_0$ describes the quasiparticle dispersion taken relative to some reference (for example,
the non-bonding energy $\varepsilon_0$). Quasiparticles complete orbits with a frequency $\omega_c$ that depends on the
scalar product \kf$\cdot$\vf\, via the expression

\begin{equation}
\omega_c(\phi,\theta)=eB{\rm cos}\theta{{\bf k}_{\rm F}(\phi)\cdot{\bf v}_{\rm F}(\phi)\over\hbar k_{\rm F}(\phi)^2}.
\label{anm}
\end{equation}
Near the bottom of the band, the quasiparticle orbits in a magnetic field appear almost circular and ${\bf v}_{\rm F}$
is isotropic and nearly parallel to the crystal momentum ${\bf k}$. As $\epsilon_{\rm F}$ approaches the van Hove
singularity (vHs) however, anisotropy in \vf\, becomes significant \cite{blundell97} and $\omega_c$ develops four-fold
anisotropy that essentially becomes infinite at the vHs.

Let us now turn to consider the analogous situation in \tl. According to recent ARPES experiments,\cite{plate05} the FS
can be fitted by a tight-binding dispersion relation

\begin{equation}
\begin{array}{c}
\varepsilon-\varepsilon_0={t_1\over2}({\rm cos}k_x + {\rm cos}k_y)+t_2({\rm cos}k_x{\rm cos}k_y)\\
+{t_3\over2}({\rm cos}2k_x + {\rm cos}2k_y)+{t_4\over2}({\rm cos}2k_x{\rm cos}k_y + {\rm cos}2k_y{\rm cos}k_x)\\
+t_5({\rm cos}2k_x{\rm cos}2k_y),
\end{array}
\label{inplane}
\end{equation}
with $t_1$ = -0.725, $t_2$ = 0.302, $t_3$ =0.0159, $t_4$ = -0.0805 and $t_5$ = 0.0034(eV).

\begin{figure}[tph]
\includegraphics[width=8cm,angle=0]{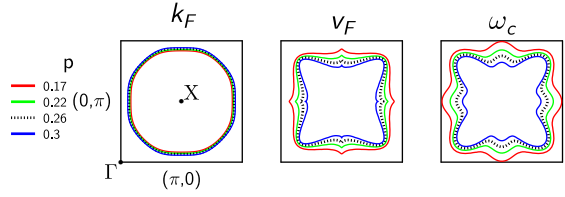}
\caption{(Color online) The parameters \kf, \vf\, and $\omega_c$ for four different doping levels ($p$ = 0.17, 0.22,
0.26 and 0.3) based on the dispersion relation given by Eq. \ref{inplane} and assuming a rigid band shift. The black
dashed line corresponds to the doping level of our $T_c$ = 17K sample.} \label{tltb}
\end{figure}

In order to visualize the doping evolution of the FS parameters according to Eq. \ref{inplane}, we show in Figure
\ref{tltb} the variation of \kf($\phi$), \vf($\phi$) and $\omega_c$($\phi$) (left, center and right panels
respectively) for different values of the chemical potential assuming a simple rigid band shift. The energy contours
have been centered on the X point of the Brillouin zone. Though the band structure may change with hole doping,
\cite{tallon} this approximation scheme serves as a good illustration of how the anisotropy in $\omega_c$ varies in a
comparable way to that in \vf. As expected, the anisotropy grows as the FS at ($\pi$, 0) approaches the vHs, though in
the doping range relevant to \tl, it never exceeds 30$\%$. Interestingly, as the chemical potential is raised, the
anisotropy of $\omega_c$ changes sign, so that the cyclotron frequency goes from being maximal to being minimal along
the zone diagonal, but retaining four-fold symmetry throughout. We approximate this without making any assumptions as
to the sign of $\omega_c$ using the expression

\begin{equation}
  \omega_c(\phi)^{-1}\approx \omega_0^{-1}(1+\beta{\rm cos}(4\phi)).
\label{waniso}
\end{equation}

The curves in the right-sided panel of Figure \ref{tltb} correspond to a range of $\beta$ values from $\beta\approx0.3$
(for $p=0.3$) to $\beta\approx 0$ at optimal doping.

The addition of this extra parameter causes only minor modifications to the fitting procedure. The conductivity is now
replaced by the equation

\begin{equation}
\begin{array}{lll}
\sigma_{zz}& =&\displaystyle {e^2\over4\pi^3\hbar^2}\int B{\rm\,cos}\theta\, dk_z^0\\
&&\\
&&\displaystyle\times\int_0^{2\pi} d\phi\int_0^\infty d\phi'{v_z(\phi,k_z^0,\varepsilon_{\rm F})\over \omega_c(\phi)} {v_z(\phi-\phi',k_z^0,\varepsilon_{\rm F})\over \omega_c(\phi-\phi')}\\
&&\\
\displaystyle&& \times{\rm exp}[h(\phi)-h(\phi-\phi')],
\end{array}
\label{condi}
\end{equation}
where $h(\phi)=-{\int{d\phi\over\omega_c(\phi)\tau_0}}$. Under isotropic circumstances $h(\phi)=-\phi/\omega_0\tau_0$ as
in Eq. \ref{condi0}. However, in the case where $\omega_c$ satisfies Eq. \ref{waniso}, this becomes

\begin{equation}
h(\phi)=-{1\over\omega_0\tau_0}\left\{\phi+{1\over 4}\beta{\rm sin}4\phi\right\}.
\label{wheq}
\end{equation}
We can now define two new periodic functions
$p_z(\phi)$ and $p_z(\phi-\phi')$ whereby

\begin{equation}
p_z(\varphi)\equiv{v_z(\varphi,k_z^0,\varepsilon_F)\over \omega_c(\varphi)}{\rm e}^{-{\beta\over4}{\rm
sin}4\varphi/\omega_0\tau_0}. \label{vi}
\end{equation}

\begin{figure*}[tph]
\includegraphics[width=15cm,angle=0]{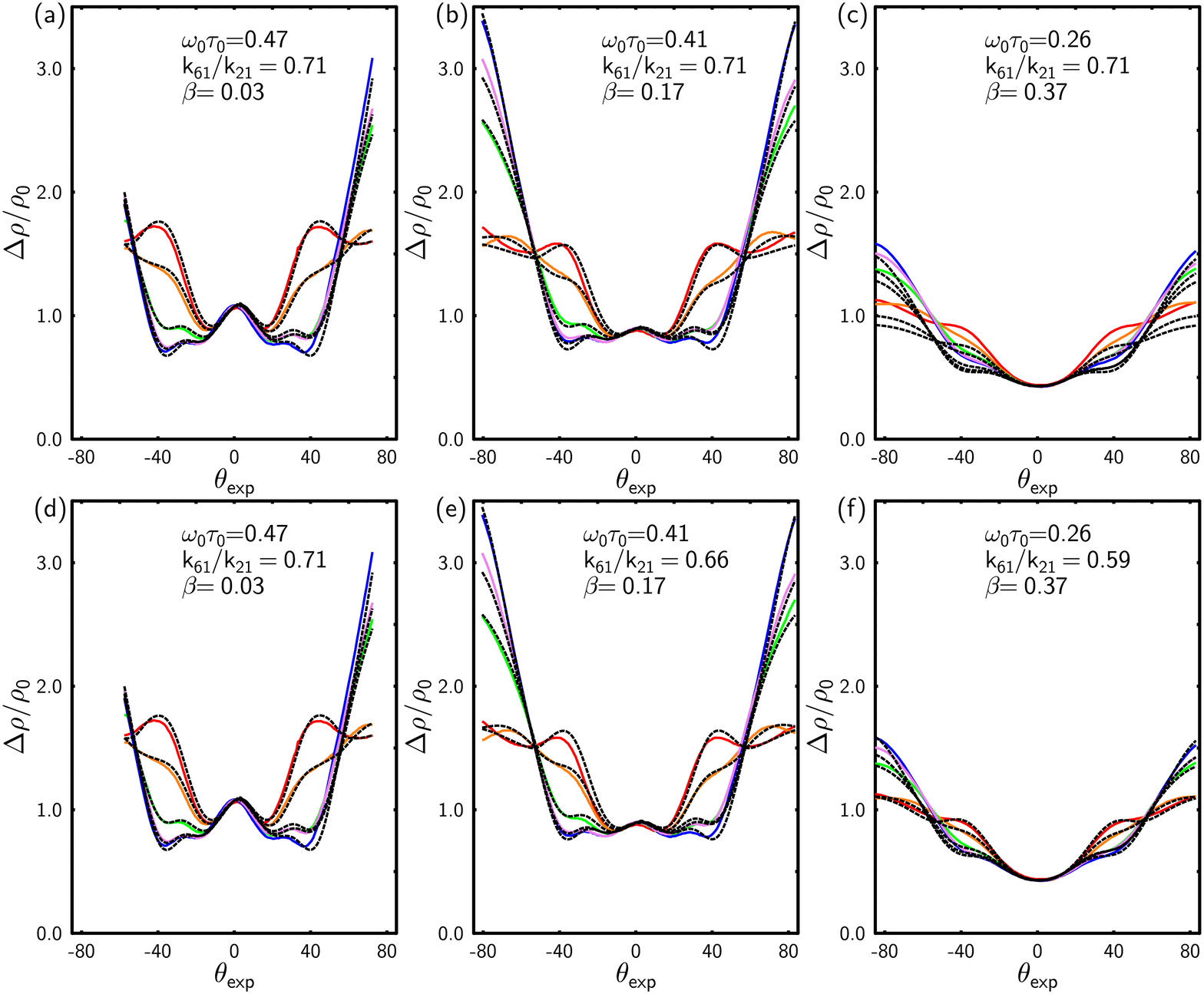}
\caption{(Color online) The raw data (solid curves with colors corresponding to $\phi_{\rm crys}$ as defined in Figure
\ref{isofit}) and best least-squares fits (black dashed lines) for ADMR taken at three different temperatures $T$ =
4.2K (panels (a) and (d)), 14K ((b) and (e)) and 50K ((c) and (f)). In panels (a), (b) and (c), the parameters
$k_{04}$, $k_{61}/k_{21}$ are fixed at their 4.2K values whilst $\omega_0\tau_0$ and $\beta$ are allowed to vary. In
panels (d), (e) and (f), $k_{61}/k_{21}$ is also allowed to vary with temperature.} \label{betacomp}
\end{figure*}

The Fourier transform of each function is given by Eq. \ref{fftvi}. The form of Eq. \ref{fftsigA} is identical, only
that $\omega_0$ is interpreted as the average of $\omega_c$ within the plane (see Eq. \ref{waniso}). The fitting
procedure proceeds as described in Section \ref{isoall} except our fitted parameters are now
$k_{04},k_{61}/k_{21},\beta$ and $\omega_0\tau_0$. As before, $k_{00}$ = 0.729 $\AA^{-1}$, $k_{04}$ = -0.022$\AA^{-1}$,
$k_{61}/k_{21}$ is fixed at low $T$ (4.2 K) and only $\omega_0\tau_0$ and $\beta$ are allowed to vary as a function of
temperature. Figure \ref{betacomp} (a)-(c) shows the best least-squares fits of the same ADMR data under this new
parameterization scheme.  While the fits are closer to the real data than in the corresponding isotropic case, there is
still a clear problem with the higher temperature fits. If we choose to allow $k_{61}/k_{21}$ to vary however, the fits
become reasonable at all temperatures, as shown in Figure \ref{betacomp} (d)-(f)). The mathematical reason for this is
that $\beta$ has two competing roles: it appears in the exponent $h(\phi)$ and in the ratio $v_z(\phi)/\omega_c(\phi)$.
In the latter, $\beta$ plays a similar role to $k_{61}/k_{21}$, as can be seen with an expansion using elementary
trigonometric identities

\begin{equation}
\begin{array}{lll}
&\displaystyle({\rm sin}2\phi+{k_{61}\over k_{21}}{\rm sin}6\phi+{k_{101}\over k_{21}}{\rm sin}10\phi)(1+\beta{\rm cos4\phi})&\\
&&\\
&=\displaystyle\left(1-\beta/2+{\beta k_{61}\over 2k_{21}}\right){\rm sin}2\phi&\\
&&\\
&\displaystyle+\left({k_{61}\over k_{21}}+\beta/2+{\beta k_{101}\over 2k_{21}}\right){\rm sin}6\phi&\\
&&\\
&\displaystyle+\left({k_{101}\over k_{21}}+{\beta k_{61}\over 2k_{21}}\right){\rm sin}10\phi+{\beta k_{101}\over 2k_{21}}{\rm sin}14\phi.&
\label{betaksixeq}
\end{array}
\end{equation}

The $k_{61}/k_{21}$, $k_{101}/k_{21}$ and $\beta$ terms can compensate each other as long as $\beta$ remains small
(that is, so long as products such as $\beta k_{101}/ 2k_{21}$ are negligible). The only non-compensating contribution
of $\beta$ in this expansion is in the multiplication of the ${\rm sin} 2\phi$ and ${\rm sin} 14\phi$ terms, though
perturbations of the former would be noticeable first. In other words, the fitting procedure tends to keep the {\it
sum} $\beta/2+k_{61}/k_{21}$ constant as a function of temperature and so the $T$-dependent changes are contained in
the behavior of $h(\phi)$. We return to this point later in our discussion of Figure \ref{betaksix}.

The changes in $k_{61}/k_{21}$ and $\beta$ required to satisfactorily fit the data are significant ($\sim$ 20$\%$
change in $k_{61}/k_{21}$ and a factor of 10 increase in $\beta$) and if correct, would imply pronounced FS
reconstruction with increasing temperature. Between 4K and 50 K, one may expect the Fermi distribution to broaden by
around 2$\%$ of the band width about the chemical potential. At $p=0.26$ this is approximately equivalent to a change
in nominal doping of $\pm 0.02$ allowing a change in $\beta$ of at most $\pm 0.05$, as is evident from Figure
\ref{tltb}. This is significantly less than is required to quantitatively account for the evolution of the ADMR. To our
knowledge, the FS restructuring required to fit the present data has never been reported in cuprates and so justifying
it would require some very subtle physical arguments. Indeed, photoemission studies have reported insignificant changes
as a function of temperature on overdoped compounds. \cite{kim02} Moreover, in a recent doping dependence study,
\cite{majed07} we found that the overall anisotropy {\it decreases} with increasing carrier concentration, i.e. as one
approaches the vHs, in marked contrast to the band structure picture discussed above and illustrated in Figure
\ref{tltb}. In the following section therefore, we turn to consider the effect of anisotropic scattering, which not
only allows the data to be fitted accurately but also avoids the physical and mathematical difficulties we have
encountered when considering anisotropy in $\omega_c$ alone.

\subsection{Anisotropic $\tau$ and anisotropic $\omega_c$}
\label{anitau}

In the most elementary description, the scattering lifetime $\tau$ is the average time between collisions of an
electron travelling in a metal. In a FL picture however, this is taken to be the mean lifetime of an electron
excitation, giving the decay time of a quasiparticle to its ground state near the chemical potential $\mu$. The rate of
change of occupation of a state at ${\bf k}$ is related to the intrinsic transition rate between two arbitrary states
\k\, and \kd, weighted by the occupation of ${\bf k}$ and the lack of occupation of state ${\bf k'}$.

The transition rate cannot be calculated without {\it a priori} knowledge of the scattering processes that are present.
In the limit of elastic scattering however, both \k\,and \kd\, are on the same energy surface, and this function is
simply cos($\phi_{kk'}$), where $\phi_{kk'}$ is the angle between \k\,and \kd. We then use the relaxation-time
approximation, whereby \kkd\,$=\mathbb{P}(\phi_{kk'})$ is a function of $\phi_{kk'}$ only and this allows a
natural definition for the scattering time $\tau$

\begin{equation}
{1\over\tau} \propto \int (1-{\rm cos}(\phi_{kk'}))\mathbb{P}(\phi_{kk'}){\rm sin}(\phi_{kk'})d\phi_{kk'}.
\label{isotau}
\end{equation}

The relaxation time approximation is often an excellent starting point for interpreting transport data and $\tau$ is
usually considered to be independent of momentum, both of the initial and final state. A more general theory however
would allow \kkd\, to be a function of the initial state \k. In this instance, it may be assumed that the form of Eq.
\ref{isotau} stays very similar,\cite{ziman61} except that \kkd\, is now \k-dependent, and hence the replacement
$\tau\rightarrow\tau({\bf k})$ needs to be made. In this case, it can be shown \cite{ziman61} that

\begin{equation}
{g_{\rm {\bf k}}\Gamma({\bf k})}\equiv{g_{\rm {\bf k}}\over\tau({\bf k})} \propto \int \{g_{\rm {\bf k}}-g_{\rm {\bf
k'}}\}\mathbb{P}_{{\bf kk'}}{\bf dk'}, \label{aniboltz}
\end{equation}

where $g_{\rm {\bf k}}=f_{\rm {\bf k}}-f^0_{\rm {\bf k}}$, $f_{\rm {\bf k}}$ and $f^0_{\rm {\bf k}}$ are the
probabilities of an electron occupying a state ${\rm {\bf k}}$ in the presence of a field and in equilibrium
respectively. $\Gamma({\rm {\bf k}})$ is the scattering rate. Eq.~\ref{aniboltz} is general enough to include inelastic
scattering mechanisms too (involving energy transfers $\leq k_BT$ for any given scattering event) and this is a direct consequence of the
relaxation-time approximation. Such details would be normally be contained in $g_{\rm {\bf k}}$, and the Boltzmann
equation would be very difficult to solve. In the relaxation-time approximation however, these details are deliberately
ignored and all that is required for Eq.~\ref{aniboltz} to hold is that the statistical ensemble of quasiparticles
returns to equilibrium between collision events. \cite{sorbello1}

Under these circumstances, we are able to define an anisotropic scattering time that will enter all of our calculations
of the conductivity. Since the scattering time {\it always} appears in the product $\omega_c\tau$ in the sum of Eq.
\ref{fftsigB} it is clear that the procedure for incorporating anisotropic $\tau$ will be identical to Section
\ref{aniomega} where we introduced anisotropy in $\omega_c$. The simplest model would involve a four-fold anisotropy
and in a similar fashion to Sandeman and Schofield\cite{sandeman01} or Ioffe and Millis \cite{ioffemillis98} we write

\begin{equation}
\Gamma(\phi) = \Gamma_0(1+\alpha{\rm cos}(4\phi)),
\label{anisotau}
\end{equation}
where $\Gamma_0\equiv1/\tau_0$.

\begin{figure}[tph]
\includegraphics[width=8cm,height=3cm,angle=0]{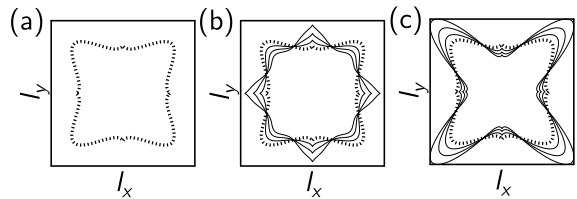}
\caption{ The mean free path \mfp\, for three different cases of the anisotropy parameter $\alpha$ for \tl\, with $p$ =
0.26: (a) the isotropic case (dashed) $\alpha=0$, (b) $\alpha=-0.1,-0.2,-0.3$ (solid, moving inwards along the zone
diagonal) and (c) $\alpha=0.1,0.2,0.3$ (solid, moving outwards along $(\pi,\pi)$).} \label{tlanitau}
\end{figure}

Figure \ref{tlanitau} illustrates the effect of this form of scattering rate anisotropy on the mean free path \mfp\,
for the FS derived in Eq. \ref{inplane} (assuming $p$ = 0.26). If $\alpha=0$, the scattering rate is isotropic and
\mfp({\bf k}) simply follows the form of \vf({\bf k}) (panel (a)). If $\alpha>0$ (panel (b), $\Gamma$({\bf k}) is
maximal in the direction parallel to the zone axes and competes with \vf({\bf k}). If, on the other hand, $\alpha<0$,
$\Gamma$({\bf k}) is maximal along the zone diagonals, the anisotropy in \mfp({\bf k}) is enhanced in this direction.

\begin{figure}[tph]
\includegraphics[width=8.5cm,angle=0]{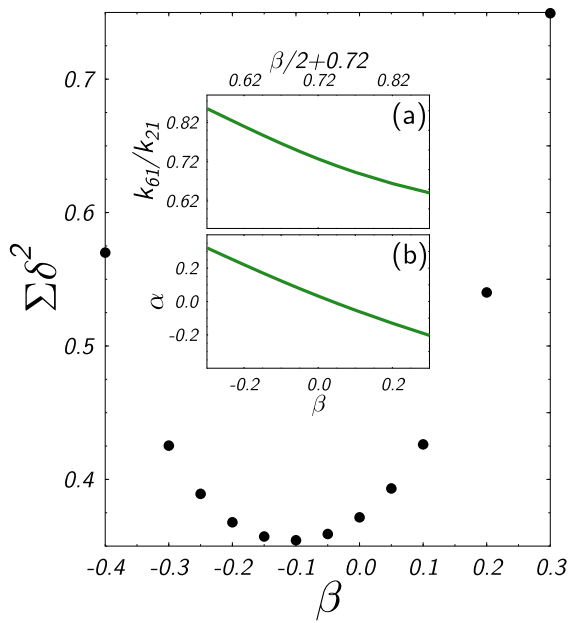}
\caption{(Color online) The quality of fit parameterised by the sum $\Sigma\delta^2$ as a function of $\beta$, which is
given a specific value between -0.4 and 0.3. Conversely, the parameters $\omega_0\tau_0,\alpha$ and $k_{61}/k_{21}$ are
free to vary. The parameter $\Sigma\delta^2$ has a broad minimum, indicating that a good fit can be achieved for a
broad range of $\beta$. Insets (a) and (b) show respectively the interdependence of $k_{61}/k_{21}$ and $\alpha$ on
$\beta$ (see also Eq. \ref{betaksixeq} and \ref{heq}). The axes in (a) are shifted as described in the text.}
\label{betaksix}
\end{figure}

With this definition of $\Gamma(\phi)$ , the conductivity is identical to Eq. \ref{condi} but now we have
$h(\phi)={\int{d\phi\over\omega_c(\phi)\tau(\phi)}}$ and

\begin{equation}
\begin{array}{l}
h(\phi)=-{1\over\omega_0\tau_0}\left\{\phi\left(1+{\alpha\beta\over2}\right)\right.\\
\left.+{1\over4}(\alpha+\beta){\rm sin}4\phi+{\alpha\beta\over16}{\rm sin}8\phi\right\},
\end{array}
\label{heq}
\end{equation}
with the periodic functions $p_z(\phi)$ and $p_z(\phi-\phi')$ now redefined as

\begin{equation} p_z(\varphi)\equiv{v_z(\varphi,k_z^0,\varepsilon_F)\over
\omega_c(\phi)}{\rm e}^{-{{(\alpha+\beta)\over4}{\rm sin}4\varphi+{\alpha\beta\over16}{\rm sin}8\varphi\over
\omega_0\tau_0}}. \label{pi}
\end{equation}

The conductivity is then given by

\begin{equation}
\begin{array}{lll}
\sigma_{zz}&=&\displaystyle{e^3\tau_0B{\rm cos}(\theta)\over 2\pi^2\hbar^2}\int dk_z^0\omega_{\alpha\beta}^{-1}\\
&&\\
&& \displaystyle\times\left[a_0c_0+{1\over2}\sum_{n=1}^{\infty}\left\{{a_nc_n+b_nd_n\over1+(\omega_{\alpha\beta}\tau_0)^2}-\right.\right.\\
&&\\
&&\displaystyle\left.\left.{(a_nd_n-b_nc_n)\omega_{\alpha\beta}\tau_0n\over1+(\omega_{\alpha\beta}\tau_0)^2}\right\}\right],
\end{array}
\label{fftsigB}
\end{equation}

where $\omega_{\alpha\beta}=\omega_0/(1+\alpha\beta/2)$. Equipped with Eq. \ref{fftsigB} we can now follow the
procedure described above. The variable parameters are now $k_{04},k_{61}/k_{21},\alpha,\beta$ and $\omega_0\tau_0$. As
before we fix $k_{00}$ (= 0.729 $\AA^{-1}$) and $k_{04}$ (= -0.022$\AA^{-1}$) and fit the low temperature data with all
other parameters free to vary. In this instance however we are now over-parameterized since $\alpha$ and $\beta$ can
compensate each other to within a factor of $\pm\alpha\beta/4$. As a consequence, one cannot accurately quote absolute
values for each parameter individually, but rather the {\it sum} $\alpha+\beta$ (see inset (b) of Figure
\ref{betaksix}).

\begin{figure*}[tp]
\includegraphics[width=15cm,angle=0]{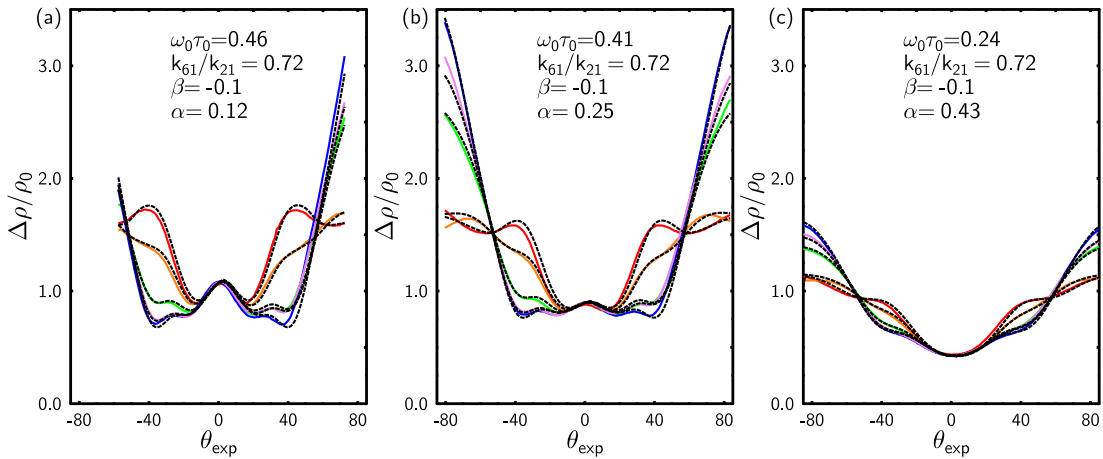}
\caption{(Color online) Raw ADMR data (solid curves with colors corresponding to $\phi_{\rm crys}$ as defined in Figure
\ref{isofit}) plotted with the best least-squares fits (black dashed lines) for three different temperatures (a) 4.2 K,
(b) 14 K, (c) 50 K. Here both $\omega_c$ and $\tau$ are considered anisotropic in the azimuthal angle $\phi$. The
parameters $k_{04},k_{61}/k_{21}$ and $\beta$ are fixed to their values determined at $T=4.2 $K whilst $\omega_0\tau_0$
and $\alpha$ are allowed to vary as a function of temperature.} \label{taucomp}
\end{figure*}


We can parameterize the quality of the fits by the sum of squared differences of the data from the fitted curve, which
is denoted $\,\Sigma\delta^2$. As $\vert\beta\vert$ becomes large, the terms $\alpha$ and $k_{61}/k_{21}$ are no longer
able to compensate and the fits decline in quality. However, as shown in Figure \ref{betaksix}, there is also a broad
flat region over which $\,\Sigma\delta^2$ is minimized and one cannot pinpoint the exact value of $\beta$. The axes in
inset (a) are shifted so that it is apparent that the sum $\beta/2+k_{61}/k_{21}$ is pinned to a value of about $0.72$,
a fact which continues to be true at higher temperatures no matter what one forces $\beta$ to be. Similarly, the value
of $\alpha+\beta$ is pinned to nearly zero at low temperature. From the low-$T$ parameterization used previously to set
the FS parameters, we can settle on a value of $\beta\approx-0.1\pm0.1$ which is comparable to that estimated from the
ARPES-derived dispersion despite being opposite in sign.

Panels (a), (b) and (c) of Figure \ref{taucomp} show the resulting fits to the new parameterization scheme, in which
only $\alpha$ and $\omega_0\tau_0$ are allowed to vary with temperature, for $T=4.2$K, $14$K and $50 $K respectively.
In contrast to previous schemes, the quality of the fits are comparable at all temperatures, without the need for any
variation in the other parameters. Hence, by introducing $T$-dependent anisotropy in the scattering rate, there is no
longer any need to invoke FS reconstruction to account for the evolution of the ADMR data. We therefore conclude that
this is the most elegant and physically realistic parameterization scheme of all those considered here.

The sign of $\alpha$ is found to be positive, indicating that scattering is weakest along the zone diagonals, as
determined previously by azimuthal ADMR measurements. \cite{hussey96} As the temperature is raised, $\alpha$ increases
markedly. This implies that the anisotropy resides in the {\it inelastic}, rather than the elastic scattering channel.
As with anisotropy in $\omega_c$, anisotropy in the scattering rate can be FS-derived, e.g. due to FS instabilities
such as charge-density waves, spin-density waves or antiferromagnetic fluctuations. Spin, charge, or indeed
superconducting fluctuations all have specific momentum (and frequency) dependence that is peaked at (or in some cases,
confined to) particular regions in {\bf k}-space. Anisotropy in $\tau^{-1}$ can also signify additional physics due,
for example, to strong electron correlations near a Mott insulating state or anisotropic electron-impurity scattering.
\cite{varmaabrahams01} The present analysis cannot of course reveal the microscopic mechanism of the anisotropic
scattering itself, but can identify some important characteristics of the scattering mechanism, such as its magnitude
or its symmetry. Systematic measurements, e.g. as a function of doping and or pressure, would then allow a detailed
comparison with the various theoretical proposals and thus help to reveal important hints as to its microscopic origin.

\section{Concluding remarks}
\label{conc}

In this paper we have set out a detailed formalism for incorporating in-plane anisotropy, both in the cyclotron
frequency and in the transport lifetime, into the analysis of interlayer magnetoresistance of a q2D metal. The focus of
the present paper has been to illustrate the need to introduce an anisotropic scattering rate in order to explain the
evolution of the ADMR data in overdoped superconducting \tl~within a Boltzmann framework. An anisotropic cyclotron
frequency $\omega_c(\phi)$ can fit the data, but only if we allow the parameters describing the Fermi surface itself to
change as a function of temperature. Given the absence of evidence for such reconstruction, this hypothesis seems
unlikely. If, on the other hand, an anisotropic scattering time is introduced, all the FS parameters can remain
constant and {\it only} $\omega_0\tau_0$ and the anisotropy in $\Gamma(\phi)$ are adjusted. Such a simple
parameterization is both elegant and experimentally accurate and we therefore believe it to be the most likely
explanation of the observed ADMR data.

A cautionary note is perhaps appropriate here. In the preceding calculations we have assumed the relaxation-time
approximation and so the microscopic relaxation dynamics have been ignored. The concept of anisotropic scattering
remains valid as long as $\tau(\phi)$ is interpreted as the exponential decay of the distribution between scattering
events \cite{sorbello1, sorbello2}. However, we cannot rule out more exotic relaxation dynamics that depend on the
presence of the magnetic field and hence may be manifested differently if probed by a different means (for example,
photoemission). We have discovered that such exotic dynamics do not need to be invoked to explain our ADMR data and
conclude that its evolution with temperature, when viewed from a Boltzmann framework using the relaxation-time
approximation, is best explained by a scattering rate with a temperature dependent anisotropy.

At high doping levels, lifetime separation in cuprates is less apparent, \cite{mackenzie96} leading some researchers to
consider the problem from this perspective. This route has the added advantage of allowing the limits of the
conventional Boltzmann transport theory to be explored as one moves across the phase diagram towards to more exotic and
potentially non-FL ground state on the underdoped side. The key message here is that by generalizing the theory to
include an anisotropic scattering rate $\Gamma({\bf k}) \equiv \tau^{-1}({\bf k})$ one can continue to apply the
Boltzmann approach and successfully account not only for the evolution of the ADMR with temperature, but also the
distinct $T$-dependencies of $\rho_{ab}$ and cot$\Theta_{\rm H}$ found in overdoped \tl. \cite{majed06} Furthermore,
initial measurements of the doping dependence of $\alpha$ in Tl2201 suggest a significant increase in anisotropy in
$\tau^{-1}({\bf k})$ as one move towards optimal doping, \cite{majed07} consistent with the observed increase in
lifetime separation (as manifest in the temperature dependence of the Hall coefficient) with decreasing doping.
\cite{kubo91, hwang94, ando04} The introduction of such anisotropy has proven a fruitful model to understand the normal
state of high-temperature superconductors, \cite{carrington92, monthouxpines92, castellani95, ioffemillis98, vdM99,
hussey03b, hussey06, dellannametzner07} though clearly more work is needed to parameterize $\tau^{-1}$({\bf k}) fully
and to identify the origin of the anisotropy.

Finally, although the focus of this paper has been a system with body-centered-tetragonal symmetry, the analysis could
very easily be generalized to layered systems of other crystallographic symmetries as already pointed out in
Ref.~\cite{kennett06} and perhaps also one-dimensional systems with anisotropic scattering.\cite{yakovenko99} ADMR
experiments on BEDT-TTF based organic superconductors have already been performed at low temperature and explained in a
Boltzmann framework without the need to invoke an anisotropic scattering rate. \cite{goddard04} A full azimuthal and
temperature dependence on other salts may however require the introduction of such a parameterization. Similarly the same ideas may
also apply to layered charge-density-wave compounds such as the rare-earth tritellurides ${\rm
RTe_3}$.\cite{analytis07} The Boltzmann equation, though simple in its assumptions, thus remains a powerful paradigm
whose explanatory power is still to be explored. ADMR is an ideal probe for just such an exploration.

We thank R. H. McKenzie, M. P. Kennett, A. Ardavan and J. A. Wilson for helpful discussions. This work is supported by
the EPSRC and a co-operative agreement between the State of Florida and the NSF. J.A. would like to thank the Lloyd's
Tercentenary Foundation.

\appendix

\section{Accounting for sample misalignment in the ADMR fitting procedure}
\label{appasym}

In this section we illustrate how sample misalignment can be accounted for. The effect of sample misalignment on ADMR
has been considered by several authors before this study, in particular Goddard \cite{goddard02} and Abdel-Jawad
.\cite{majedthesis} Before we begin, let us define two frames of reference, that of the laboratory $x_p, y_p, z_p$, in
which the field is parallel to the $z_p$ direction, and that of the crystal $x_c, y_c, z_c$. The $y_p$ axis is taken as
the axis of polar rotation and $x_p$ is perpendicular to this. The normal to the crystallographic plane shall be
defined as $z_c$, and the in-plane directions $x_c$ and $y_c$ shall be taken to be parallel and perpendicular to the
copper-oxide bonds respectively. The angle between the field direction and the crystallographic normal $z_c$ gives the
crystallographic polar angle $\theta$. The projection of the field onto the $x_c-y_c$ plane gives the crystallographic
azimuthal angle $\phi$, taken from the $x_c$ axis.

\begin{figure*}[tph]
\includegraphics[width=13.2cm]{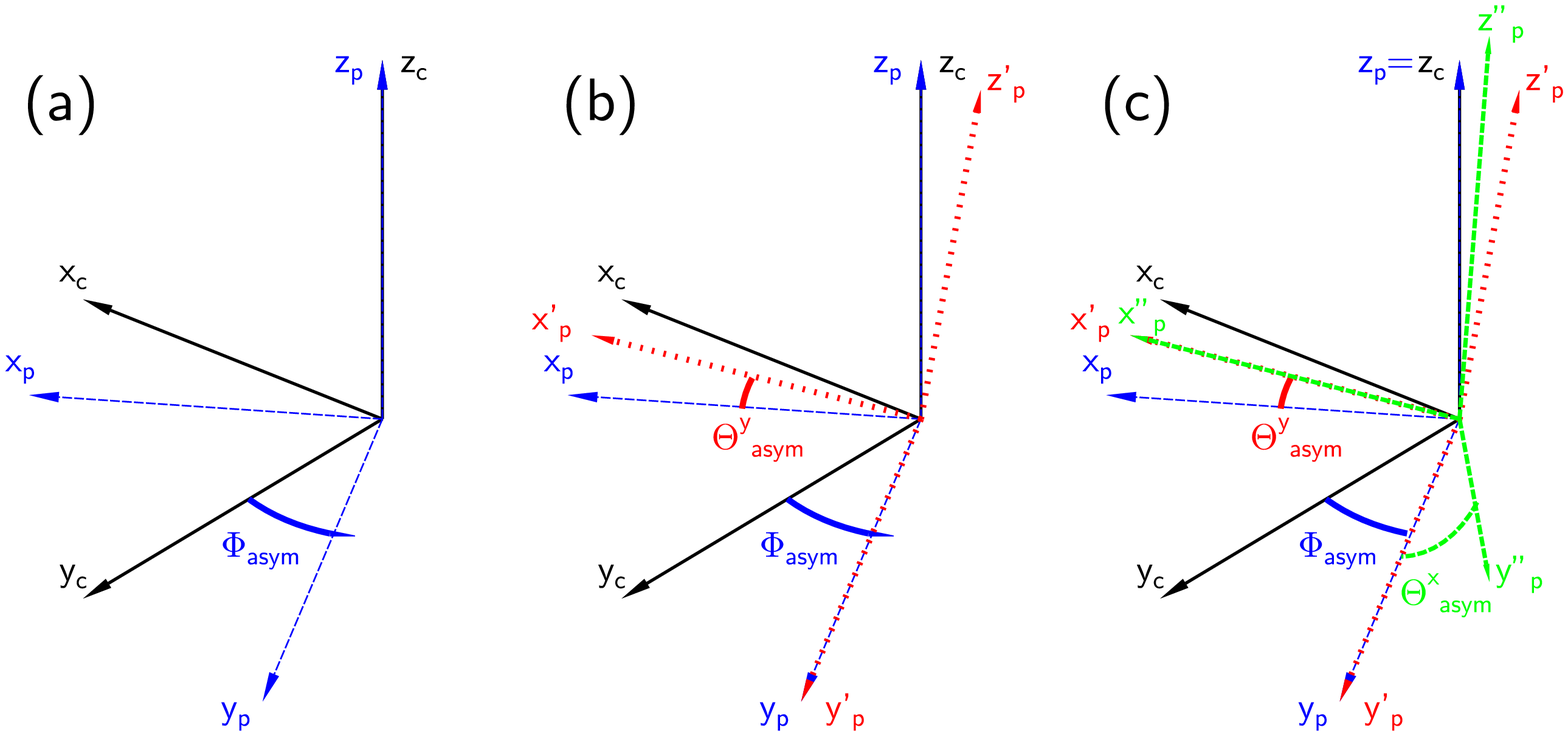}
\caption{(Color online) The reference frame of the laboratory $x_p, y_p, z_p$ and that of the crystal $x_c, y_c, z_c$
where $x_c$ is parallel to the copper-oxide bonds. (a) The first misalignment considered is that where the experimental
$z_p$ and crystal $z_c$ directions coincide but the axes on the azimuthal plane are offset by an amount $\Phi_{\rm
asym}$. The second misalignment considered is that where the experimental and crystal $z$ directions do not coincide.
Two rotations are responsible, one about the platform $y_p$ axis by an angle $\Theta_{\rm asym}^y$ shown in (b), and
another about $x_p'$ by an angle $\Theta_{\rm asym}^x$ shown in (c). This gives completely general description of the
crystal misalignment with respect to the platform axes.} \label{psias}
\end{figure*}

There are two important differences between this study and that of Goddard {\it et al.}. \cite{goddard02} Firstly,
instead of correcting experimental $\theta_{\rm exp}$ and $\phi_{\rm exp}$ for misalignment to find the appropriate
crystallographic $\theta$ and $\phi$, we fit the experimental data by including the misalignment in the fitting
procedure.  Secondly, in the analysis of Goddard,\cite{goddard02} the crystallographic axes $x_c$ and $y_c$ can fall
anywhere in the plane of the crystal and do not have assigned direction with respect to the crystal bonds. This gives
the misalignment one less parameter, and one of the crystal axes can always fall somewhere in the $x_p-y_p$ plane of
the laboratory frame. In the present analysis this is not the case and so three rotations need to be included in the
fitting procedure in order to account for every possible misalignment:

\begin{itemize}

\item Beginning with the sample aligned with the laboratory frame,there is a misalignment in azimuthal angle denoted by $\Phi_{\rm asym}$,
rotated about $z_p$ as shown in Figure \ref{psias} (a).

\item A rotation about the $y_p$ axis denoted by $\Theta_{\rm asym}^y$ as shown in Figure \ref{psias} (b).

\item A rotation about the $x_p$ axis denoted by $\Theta_{\rm asym}^x$ as shown in Figure \ref{psias} (c).

\end{itemize}

\begin{figure}[tph]
\includegraphics[width=4.5cm]{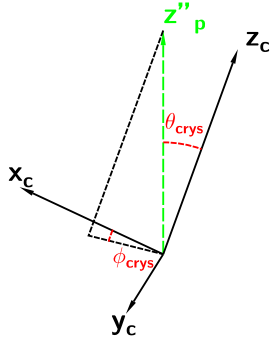}
\caption{(Color online) The polar $\theta_{\rm crys}\equiv\theta$ and azimuthal $\phi_{\rm crys}$ angles that enter Eq.
\ref{condi} depend only on the projection of $z_p''$ on the crystal axes. This projection will change as a function of
$\theta_{\rm exp}$ and $\phi_{\rm exp}$ and so must be calculated at each point.} \label{thetacrys}
\end{figure}

These transformations are elegantly described algebraically. We follow the notation whereby a rotation
$R_{{\bf\beta}}(\theta){\bf \alpha}$ is a rotation of a vector ${\mathbf \alpha}$ by angle $\theta$ about an axis
${\mathbf \beta}$. In particular the laboratory axis $z''_p$ is tranformed relative to the crystallographic $z_c(0,0)$
(before any azimuthal or polar rotation) axis to

\begin{equation}
z''_p = R_{x'_p}(\Theta^x_{\rm asym})R_{y'_p}(\Theta^y_{\rm asym})z_c(0,0),
\end{equation}
due to the misalignment of the sample. In an ADMR experiment, the sample is then rotated about the laboratory $z''_p$
azimuthally by an angle $\phi_{\rm exp}$, and then rotated about the $x''_p$ axis a polar angle $\theta_{\rm exp}$. The
position of the crystallographic $z_c(\theta_{\rm exp},\phi_{\rm exp})$ after these azimuthal and polar rotations axis
is given by

\begin{equation}
z_c(\theta_{\rm exp},\phi_{\rm exp}) = R_{x''_p}(\theta_{\rm exp})R_{z''_p}(\phi_{\rm exp})z_c.
\end{equation}
With reference to Figure \ref{thetacrys} it is elementary to see that the projection of the field parallel to $z''_p$
on the crystallographic $z_c(\theta_{\rm exp},\phi_{\rm exp})$ will give $\theta_{\rm crys}$, which should be used to
calculate the value of the magnetoresistance in the analysis. Similarly, the projection on the $x_c(\theta_{\rm
exp},\phi_{\rm exp})-y_c(\theta_{\rm crys},\phi_{\rm exp})$ plane will yield $\phi_{\rm crys}$. Algebraically we have,

\begin{equation}
{\rm cos}(\theta_{\rm crys}) = z''_p \cdot z_c(\theta_{\rm exp},\phi_{\rm exp}),
\end{equation}
and

\begin{equation}
{\rm tan}(\phi_{\rm crys}) = {z''_p \cdot x_c(\theta_{\rm exp},\phi_{\rm exp})\over z''_p \cdot y_c(\theta_{\rm
exp},\phi_{\rm exp})}.
\end{equation}
The asymmetries for the sample considered in the present paper were estimated to be $\Phi_{\rm asym}=8^\circ$,
$\Theta_{\rm asym}^y=0^\circ$, $\Theta_{\rm asym}^x=3^\circ$. The parameter $\Phi_{\rm asym}$ was approximately equal to values
estimated from diffractometry performed after the ADMR experiment.


\begin{thebibliography}{1}
\bibitem{asym} Strictly speaking, $\phi_{\rm crys}$ and $\theta_{\rm crys}$ are equal to $\phi_{\rm exp}$ and
$\theta_{\rm exp}$ respectively only when the crystal axes are perfectly aligned with the platform axes. As this is
very difficult to achieve, the experiment gives only $\phi_{\rm exp}$ and $\theta_{\rm exp}$ and the angle with respect
to the crystal needs to be calculated by fitting the misalignment angles ${\rm\Phi_{\rm asym},\Theta_{\rm
asym}^x,\Theta_{\rm asym}^y}$(as described in Appendix \ref{appasym}).
\end{thebibliography}
\end{document}